\documentclass[12pt,manuscript]{aastex}

\newcommand{\mic}{~$\mu$m}

\shorttitle{Deep Extinction Law}
\shortauthors{Rom\'an-Z\'u\~niga et al.}

\begin{document}

\title{The Infrared Extinction Law at Extreme Depth\\
in a Dark Cloud Core}

\author{Carlos G. Rom\'an-Z\'u\~niga}
\affil{Harvard Smithsonian Center for Astrophysics, Cambridge, MA 02138}
\email{cromanzu@cfa.harvard.edu}

\author{Charles J. Lada}
\affil{Harvard Smithsonian Center for Astrophysics, Cambridge, MA 02138}
\email{clada@cfa.harvard.edu}

\author{August Muench}
\affil{Harvard Smithsonian Center for Astrophysics, Cambridge, MA 02138}
\email{gmuench@cfa.harvard.edu}

\and

\author{Joao Alves}
\affil{Centro Astrof\'isico Hispano Alem\'an, Granada, Spain}
\email{jalves@caha.es}

\begin{abstract}
%We combined sensitive near-infrared data obtained with ground-based imagers on the ESO NTT and VLT telescopes with space mid-infrared data acquired with the IRAC imager on the Spitzer Space Telescope to calculate the extinction law $A_\lambda/A_{K_S}$ as a function of $\lambda$ between 1.25 and 7.76\mic~to an unprecedented depth in \object{Barnard 59}, a star forming, dense core located in the Pipe Nebula. The ratios $A_\lambda/A_{K_S}$ were calculated from the slopes of the distributions of sources in color-color diagrams $\lambda-K_S$ vs. $H-K_S$. The distributions in the color-color diagrams are fit well with single slopes to extinction levels of $A_{K_S}\approx 7$ ($A_V\approx 59$ mag). Consequently, there appears to be no significant variation of the extinction law with depth through the B59 line of sight. However, when slopes are translated into the relative extinction coefficients $A_\lambda/A_{K_S}$, we find an extinction law which is slightly flatter than a dust extinction model for a cloud with a total to selective absorption $R_V$=5.5, suggesting increased absorption compared to that expected for diffuse clouds. This difference is possibly due to the effect of grain growth in denser regions. Finally, our slopes in our diagrams are systematically less steep than those from the study of \citet{idb05} for clouds with lower column densities, and this indicates that the extinction law between 3 and 8\mic~might vary from place to place in the Galaxy.

We combined sensitive near-infrared data obtained with ground-based imagers on the ESO NTT and VLT telescopes with space mid-infrared data acquired with the IRAC imager on the Spitzer Space Telescope to calculate the extinction law $A_\lambda/A_{K_S}$ as a function of $\lambda$ between 1.25 and 7.76\mic~to an unprecedented depth in \object{Barnard 59}, a star forming, dense core located in the Pipe Nebula. The ratios $A_\lambda/A_{K_S}$ were calculated from the slopes of the distributions of sources in color-color diagrams $\lambda-K_S$ vs. $H-K_S$. The distributions in the color-color diagrams are fit well with single slopes to extinction levels of $A_{K_S}\approx 7$ ($A_V\approx 59$ mag). Consequently, there appears to be no significant variation of the extinction law with depth through the B59 line of sight. However, when slopes are translated into the relative extinction coefficients $A_\lambda/A_{K_S}$, we find an extinction law which departs from the simple extrapolation of the near-infrared power law extinction curve, and agrees more closely with a dust extinction model for a cloud with a total to selective absorption $R_V$=5.5 and a grain size distribution favoring larger grains than those in the diffuse ISM. Thus, the difference we observe could be possibly due to the effect of grain growth in denser regions. Finally, the slopes in our diagrams are somewhat less steep than those from the study of \citet{idb05} for clouds with lower column densities, and this indicates that the extinction law between 3 and 8\mic~might vary slightly as a function of environment.

\end{abstract}



\section{Introduction}

One common technique to study the structure of molecular clouds is to map their distribution of dust extinction (which in turn is
proportional to the column density of $H_2$) by measuring the reddening of background stars. This reddening is 
observable as an excess in a given photometric color, which is translated to extinction by assuming a conversion factor from a previously determined extinction law \citep[e.g.][]{NICE,NICER}. Towards many molecular clouds dust extinction is so large that reddened background stars are only detectable in the infrared, making it necessary to image clouds with specific detectors for that regime. Under the best conditions (high data quality, high density of background sources based on the location of the field against the Galactic plane, etc.), it is possible to use near-infrared (NIR)\footnote{In this paper we define Near-infrared (NIR) as the range from 1.2 to 2.2\mic~(J,H,K) typically covered from ground, and Mid-infrared (MIR) as the range from 3.5 to 8.0\mic~covered by IRAC.} cameras on large aperture telescopes (8m class) to observe colors of stars in the background of molecular cores with maximum extinctions reaching up to $A_V$=50 mag \citep[e.g.][]{joao_origins}. Higher extinction regimes (50$<A_V<$100 mag), such as those found in some very dense, protostellar cores and more recently in Infrared Dark Clouds (IRDCs) \citep[e.g.][]{teyssier02}, can not be probed even with those instruments and the best observational conditions. If one wishes to measure extinction at those levels it is necessary to consider the use of MIR wavelengths, which are capable of penetrating deeper into the cores.

The Infrared Array Camera (IRAC) on board the Spitzer Space Telescope, with its relatively large field of view ($5.1\times5.1\arcmin$) and its capability of simultaneous observation of four MIR bands ([3.6], [4.5], [5.8] and [8.0]\mic) allows us to detect those deeply extincted sources that are otherwise hidden in the optical and NIR, and to measure with reasonable precision color excesses that correspond to the highest extinction levels of the most opaque pre-stellar and young star forming cores. Combined NIR+MIR colors of the form $\lambda-K$ can be used to determine the color excesses of those sources with no $J$ (1.2\mic) or $H$ (1.6\mic) detection, while stand alone MIR colors can be used for sources hidden even at $K$ (2.2\mic). The conversion of color excesses to extinction for deeply embedded sources, however, requires of knowledge of the extinction law in the NIR and MIR.

In the NIR regime, the extinction law has been determined in numerous studies \citep[e.g.][]{rile85,kbl96,lal06}, and it has been shown to be an approximately uniform power law of the type $\lambda^\beta$, with $\beta \approx -1.8$. The NIR extinction law appears to be universal and independent of $R_V$, the total to selective extinction ratio, although slight variations from place to place and as a function of depth (i.e. extinction level) have been reported \citep{moore05,nishi06}.
 
In the MIR regime, the determination of the extinction law has been more difficult because this wavelength range is best accessed from space. Spectroscopic studies with the SWS spectrometer on the Infrared Space Observatory (ISO) have been used to determine the extinction law by comparing observed and expected intensities at specific MIR wavelengths for different dust extinction models. In one of these studies, \citet[][hereafter L96]{lutz96} obtained the 2.5-45\mic~spectrum of SgrA*, and their observed extinction law was found to be almost flat between 4 and 8\mic, instead of continuing the NIR power law towards longer wavelengths, as expected from models of clouds with total to selective extinction $R_V$=3.1 (typical of the diffuse ISM) and a standard mixture of graphite-silicate grains \citep[e.g.][]{lidra01}. The extinction law of L96  instead displayed a better agreement with models which are constructed with both a smaller number of very small grains and a more significant contribution of large grains as might be expected for more dense clouds ($R_V>4.0$) \citep[see][hereafter WD01]{wd01}. However, the spectroscopic studies by \citet{bert99} and \citet{rosen00} produced significantly different results: they calculated relative line intensities of H$_2$ emission towards the Orion Molecular Cloud (OMC) also from SWS spectra, but their measurements are in better agreement with an extinction law that extends the power-law observed in the NIR, consistent with diffuse medium models.

\Citet[][hereafter I05]{idb05} combined IRAC and 2MASS for two distinct fields of the Galactic plane to determine the MIR extinction law photometrically from color excesses of background stars as done in the NIR. They derived the extinction law in the form $A_\lambda/A_K$ as a function of $\lambda$ between 1.2 and 8.0\mic~. The extinction law was found to be very similar in both fields, and in closer agreement with the observations of L96 and therefore with models of denser clouds with larger grains. The studies of \citeauthor{jiang03} (\citeyear{jiang03,jiang06}) used data from the ISOGAL survey to estimate the extinction at 7\mic~relative to $K$, and their data showed a good agreement with I05 in a relatively large number of fields across the Galactic plane, perhaps suggesting a universal MIR extinction law for clouds across the Galactic plane.

However, one very important aspect to consider is that the regions of the Galactic plane studied to date have only moderate extinctions. In particular, the fields studied by I05 scarcely contained material with very large column densities: one of their fields points towards $l=42\deg$, which despite coinciding to some extent with with clouds in the Galactic Ring, W50 and the Sagittarius Arm, is mostly high above the plane and thus covers probably more diffuse material. The other line of sight of I05 crosses the Carina Arm, and coincides with a region dense enough to have ongoing star formation, but given the limited sensitivity of the 2MASS catalog, they could not obtain reliable colors beyond $H-K=$1.2, and thus their study is limited to zones with extinctions below $A_K\lesssim$2 mag. It would be important now to determine if the extinction law remains universal toward regions at much higher extinction levels ($2\lesssim A_K\lesssim 10$), like those that characterize molecular cloud cores that form stars. At such regions, extinction can easily exceed the depth probed by I05, and the properties of dust might vary significantly because of growth and coagulation of grains.

In order to test the universality of the extinction law at extreme opacity, we combined deep, high resolution NIR ($J$, $H$ and $K_S$) observations from ESO-SOFI/NTT and ISAAC/VLT with recently available Spitzer Space Telescope observations in the four IRAC MIR bands ([3.6], [4.5], [5.8] and [8.0]\mic) of the dark core \object{Barnard 59} (B59, LDN1746), which is the most opaque region of the Pipe Nebula ($l$=2.9,~$b$=7.12,~$d_\odot$=130 pc) and is currently forming a star cluster \citep{brooke06}. B59 also has the advantage of being projected against the highly uniform background of the Galactic bulge, which guarantees that any variations of the intrinsic colors of background sources will be insignificant.  Our data allowed us to measure color excesses for numerous sample of deeply extincted stars behind the core otherwise hidden to NIR observations alone. By using the color excess of such sources we are able to extend the reddening law between 1.2 to 8.0\mic~to unprecedented regimes of extinction.

\section{Observations}

\begin{figure}
\plotone{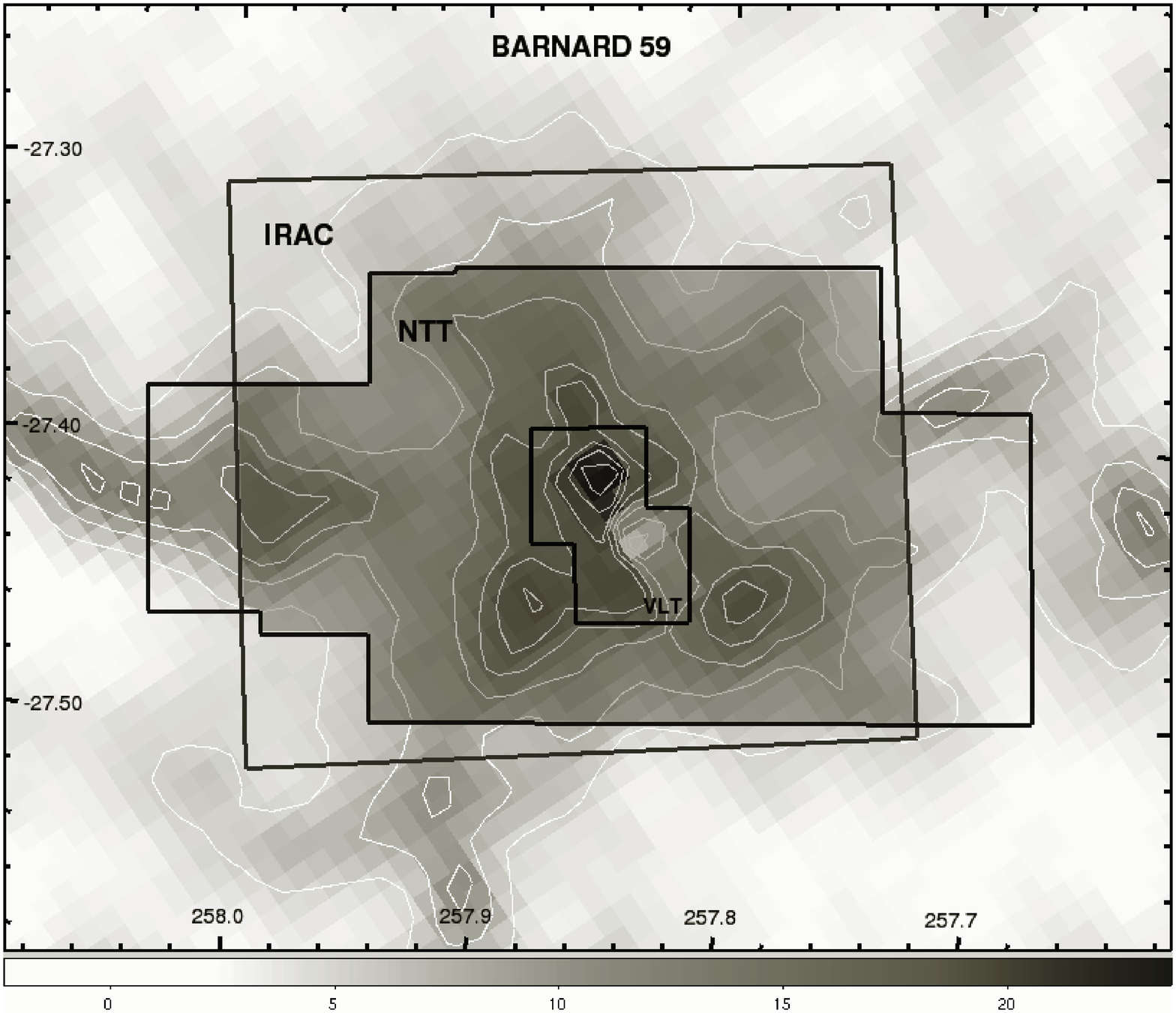}
\caption{Common observations of B59. The large trapezoidal area denotes the coverage of IRAC. The irregularly shaped medium size area delineates the span of the NTT observations and the small region at the center indicates the fields observed with the VLT. The common, shaded region is the one used in this study. The halftone and contours in the background indicate extinction $A_V$ from 5.0 to 25.0 mag, in steps of 2.5 mag, as determined by \citet{lal06}.}
\label{fig:b59regions}
\end{figure}

We selected ten SOFI/NTT and two ISAAC/VLT fields from our large scale NIR survey of the Pipe Nebula, which is part of a multi-wavelength project to study the primordial conditions of star formation in this nearby, very young molecular cloud. The selected fields cover the regions of highest extinction for B59 and its surroundings, as determined from the 2MASS extinction map of \cite{lal06}. In Figure \ref{fig:b59regions} we plot the location and extent of the NTT and VLT observations along with the coverage of the IRAC mosaic to show the area of coincidence of the surveys. All SOFI/NTT fields were observed in $H$ and $K_S$, and four of them were also observed in $J$. The ISAAC/VLT fields were only observed in $H$ and $K_S$.

The observations at NTT were carried out during the month of June in 2002 and 2003, while the VLT observations were done in July, 2003. Raw data were processed using standard IRAF pipelines, which were modified from those used in the NOAO/FLAMINGOS survey of Giant Molecular Clouds. One pipeline \citep[see][]{mythesis} processes all raw frames by subtracting darks and dividing by flat fields, optimizing signal to noise ratios by means of a two pass sky subtraction method, and combining reduced frames. The product images from the first pipeline were analyzed by a second pipeline \citep[see][]{levinethesis}, which extracts all possible sources from a given field, applies PSF photometry, calibrates to a zero point and calculates an accurate astrometric solution. Both photometry and astrometry were calibrated with respect to 2MASS.

The resultant photometry for the SOFI/NTT fields is reliable ($\sigma_{J,H,K_S}<$0.1 mag) down to $J$=20.4, $H$=19.2 and $K$=18.6 mag. The ISAAC/VLT observations, aimed to increase as much as possible the detection of sources near the core center, yield reliable photometry down to $H$=22.2 and $K_S$=20.9. 

 In the case of the Spitzer observations, we made use of the publicly available pipeline catalog of B59 which is part of the Cores to Disks Legacy Project (Data Release 3, S11.0) \citep{C2d} and includes photometry in the four IRAC bands, [3.6], [4.5], [5.8] and [8.0].

\section{Analysis}

The relative extinction curve, $A_\lambda / A_{K_S}$ ($\lambda$=[3.6] to [8.0]) for B59 was constructed following a similar technique to the one used by I05, except that we calculated the relevant color ratios with respect to $H-K_S$ instead of $J-K_S$. We assumed a fixed ratio $A_H/A_K=1.55$, which corresponds to the one determined by I05 and agrees well with the value of 1.56 obtained by \citet{rile85}. This choice serves as a simple normalization of our law to the one of I05, and any variations in our determinations will be relative to their work. With this value, the relative extinction is calculated as:

\begin{equation}
{A_\lambda \over A_{K_S}} = ({{A_H \over A_{K_S}} -1})\times {E_{\lambda-K_S}\over E_{H-K_S}}+1.
\end{equation}

The quotient $E_{\lambda-K_S} / E_{H-K_S}$ is simply the slope of the distribution of sources in the color-color diagram ($\lambda-K_S$ vs. $H-K_S$). The slopes were calculated with a weighted least square fit applied to restricted color samples. A first restriction limited the allowed photometric uncertainties to 0.1 mag in each NIR and MIR band. A second restriction assured that only sources with no intrinsic infrared excesses were considered, by separating out from the combined catalog every star that had a SED indicative of intrinsic excess (i.e. YSOs and galaxies), and keeping only those sources that fall in the `star' or 'two band' object type categories in the IRAC catalog \citep[see][table 8]{C2d}. 

As a result of these restrictions, the total number of sources used to calculate the linear fits in the $J-K$ vs. $H-K_S$, $[3.6]-K_S$ vs. $H-K_S$, $[4.5]-K_S$ vs. $H-K_S$, $[5.8]-K_S$ vs. $H-K_S$ and $[8.0]-K_S$ vs. $H-K_S$ diagrams are 672, 2779, 2776, 1701 and 1170 respectively. From now on, we will use the nomenclature $R[J]$, $R[3.6]$, $R[4.5]$, $R[5.8]$ and $R[8.0]$ when referring to the color-color diagrams constructed with these samples. 

The least square fits were calculated with a technique similar to that of \citet{lal06}: First we divided the $H-K_S$ data set into bins of equal size. The bins have widths of 0.056 in $R[J]$, 0.11 mag in $R[3.6]$ and $R[4.5]$, and 0.14 mag in $R[5.8]$ and $R[8.0]$. These bin widths are equivalent to dividing the $H-K_S$ data set in bins of constant `zero excess' extinction\footnote{The `zero excess' extinction is defined as $A_{K_S,0}=1.78*(H-K_S)$, which assumes an intrinsic value $(H-K)_0=0.0$.}: 0.1 mag for $R[J]$, 0.2 mag for $R[3.6]$ and $R[4.5]$, and 0.25 mag for $R[5.8]$ and $R[8.0]$. These bin sizes were set by the requirement that a minimum number of sources are included in each bin to assure more robust statistics. Then we calculated the median and the standard deviation of the two corresponding colors in each bin and applied a least squares linear fit with chi-square minimization to the fiducial defined by these median points; lastly, we iterated the process by recalculating the fit with only those objects that are within 5$\sigma$ of the median points, until the value of the slope converges within $10^{-4}$. 

In Figures \ref{fig:fit1} and \ref{fig:fits2to5} we show the linear fits applied to samples in the diagrams $R[J]$ to $R[8.0]$. As mentioned above, the slope of each distribution is the corresponding color excess ratio. However, as an independent check we also calculated the color excess ratios by applying a method similar to the one used by \citet{kbl96} (hereafter KBL), in which a set of color excesses for each source was calculated by individually subtracting from it the colors of each star in a control field. 

Our control field is located approximately two degrees West of Barnard 59 and at the same galactic latitude. We also have NIR ESO observations as well as C2D catalog for this field. The measured photometric dispersion for sources in these control areas is less than 0.08 mag in all the $\lambda-K_S$ colors, which is comparable to the median photometric uncertainty of the colors, showing that the bulge field in the background of the Pipe Nebula has such a small intrinsic color dispersion that our photometry cannot even resolve it. The resultant color-excess data sets are plotted against each other and again, the slopes estimated from the least square fits gave directly the ratios of color excess. Comparison of the results from the two fitting methods was particularly useful for the $R[J]$ diagram, because the number of $J$ detections was small compared to those in the other bands. In the case of the color-color method some bins had very few sources and therefore poor statistics, which resulted in a least square fit which was also poor, with a slope uncertainty of 0.18. When the color-excess method was applied, the statistics per bin became more robust and the uncertainty in the slope was reduced to only 0.05. The slopes calculated for the source distributions on the color-color and color-excess diagrams, agree within the error in all cases, $R[J]$ to $R[8.0]$.

\section{Results and Discussion}

In Table \ref{tab:extlaw} we list the slopes of the linear fits from the color-color and color-excess diagrams for $R[J]$ to $R[8.0]$. For comparison, we also list the corresponding slopes for the average data points of I05, inferred from the factors $A_\lambda/A_K$ quoted on their paper. The table also presents the relative extinction ratios $A_\lambda /A_{K_S}$, which define the extinction law determined in this study. The uncertainties listed are those calculated from the uncertainties in the slopes of our color-color and color-excess diagrams, and do not include the uncertainty of 0.08 in the ratio $A_H/A_{K_S}=1.55$ of I05.  

Our distributions, just as those of I05, are well fit with a single slope: we do not see any clear evidence for a variation in the slope of the source distributions with increasing extinction, even though the background sources we observe through Barnard 59 are much more obscured ($A_V\lesssim 59$, $A_K\lesssim 7$ mag) than those studied by I05 and other authors (e.g. $A_V\lesssim17$, $A_K\lesssim2$ in I05). 

\begin{figure}
\includegraphics[angle=90,scale=0.75]{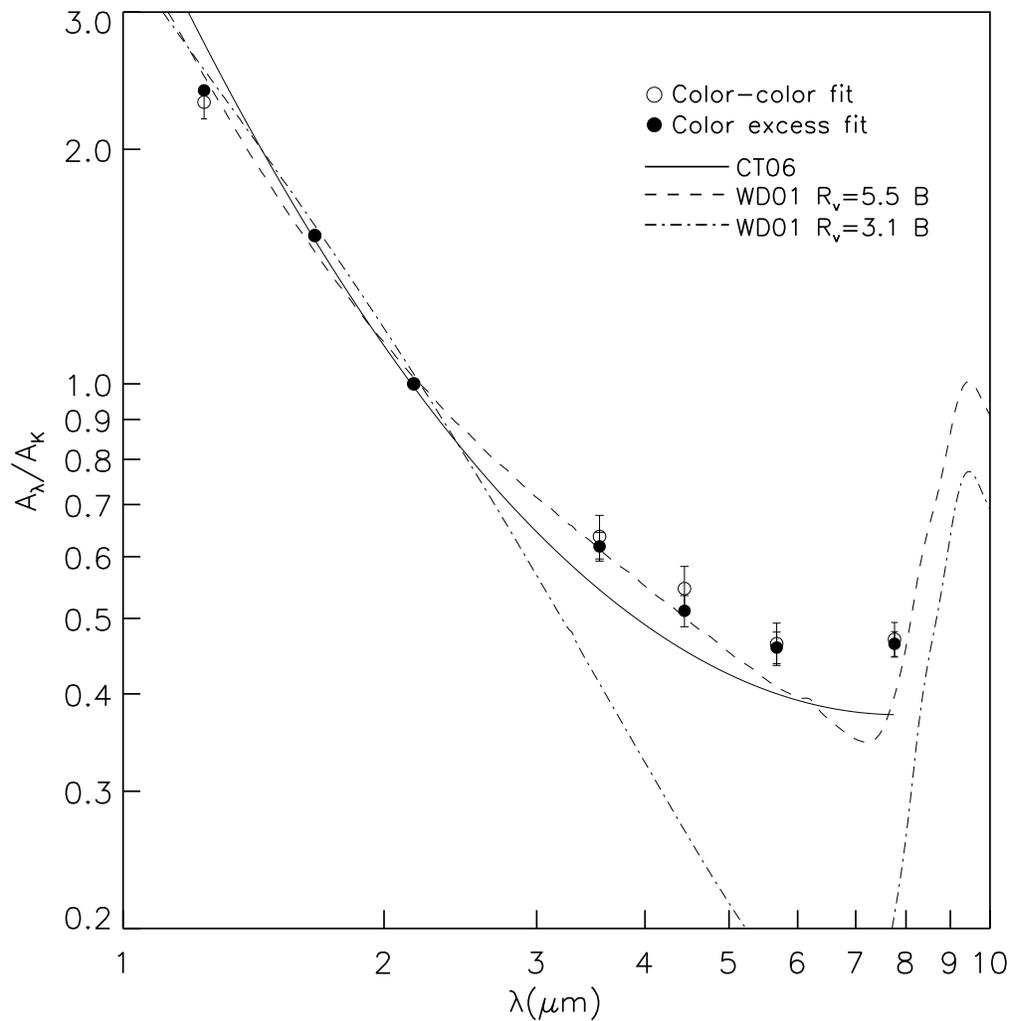}
\caption{The extinction in Barnard 59 as determined with our analysis. Open circles represent the relative extinction ratios $A_\lambda/A_{K_S}$ calculated directly from the fits to points in color-color diagrams. Solid circles symbols are extinction ratios calculated from the fits to the distributions of color excesses (KBL method). The polynomial fit to data points of I05 and L96 by \citet{ct06} (CT06) is shown with a solid line. The `case B' synthetic extinction curves of WD01 for $R_V$=3.1 and 5.5 are plotted with dot-dashed and dashed lines respectively.}
\label{fig:relative}
\end{figure}

Our study confirms that the extinction law in dense molecular clouds is approximately `flat' between 3 and 8\mic. In Figure \ref{fig:relative} we show how our extinction law compares to the `case B' synthetic extinction curves of \citet[][hereafter WD01]{wd01} for $R_V$=3.1 (diffuse cloud) and $R_V$=5.5 (dense cloud). The case B model of WD01 keeps a constant abundance of carbon and silicates, resulting in a distribution of grain sizes that favors fewer small grains and a more significant contribution of large grains in dense cloud dust, as suggested by observations \citep[e.g.][]{kimartin96}. The $R_V$=5.5 curve is in much closer agreement with our data and those of I05 and L96 than the $R_V$=3.1 curve. In general terms, our data confirm that towards dense molecular cloud material, the MIR extinction law departs from the simple extrapolation of the NIR power-law to longer wavelengths. The observed opacity in dense clouds appears to become higher and more gray in the MIR, with a soft inflexion towards the silicate peak at 10\mic.

An extinction law that departs from the NIR power-law in such manner could be showing how absorption by dust might become more gray in a very dense cloud due to anticipated grain growth at high column density regimes. In current models \citep[e.g.][]{ligreen03}, as a molecular cloud becomes more dense (e.g. when evolving towards protostellar collapse), grains are expected to grow via coagulation of smaller grains and via the accretion of an organic refractory mantle composed of ices of species depleted from the gas phase. In the first case, there is a net removal of small grains ($0.1$\mic) and PAHs in the dense cloud \citep{chokshi}, which results in a significant contribution to extinction by particles with larger cross sections. In the second case, the contribution of an accreted ice mantle to the total extinction at any given wavelength is expected to be less important because the mantles may not increase the grain sizes by more than about 18\% \citep{dra85}. In any event, although our observations may suggest that the dust grains in a dense cloud are larger than those expected for diffuse gas, we find no significant evidence for additional grain growth as a function of increasing extinction, and presumably density, within B59 itself.

The extinction law we determined towards B59 appears to be different from the one inferred from the SWS observations of the Orion Molecular Cloud by \citet{bert99} and \citet{rosen00}, which is similar to the one expected from models of diffuse clouds \citep[e.g.][]{lidra01}. However it is not clear that the OMC is a good example of a diffuse cloud: even though the Orion cloud is located at a high galactic latitude, meaning a very small contamination by foreground dense clouds, one has to consider that much of the dusty material observed toward the OMC probably arises in the HII region and in dense clouds behind it. Indeed, as pointed out by \citet{bert99} their ``effective extinction may not necessarily follow an average interstellar extinction law, since the emitting and absorbing gas may be mixed''. In addition, \citet{rosen00} warned that there are large uncertainties in their calculations, arising mostly from their calibration and the presence of foreground $H_2$ fluorescence, but that the minimum of opacity at 6.5\mic~(which implies the continuation of the NIR power law) they observe could still be real. 

\begin{figure}
\includegraphics[angle=90,scale=0.51]{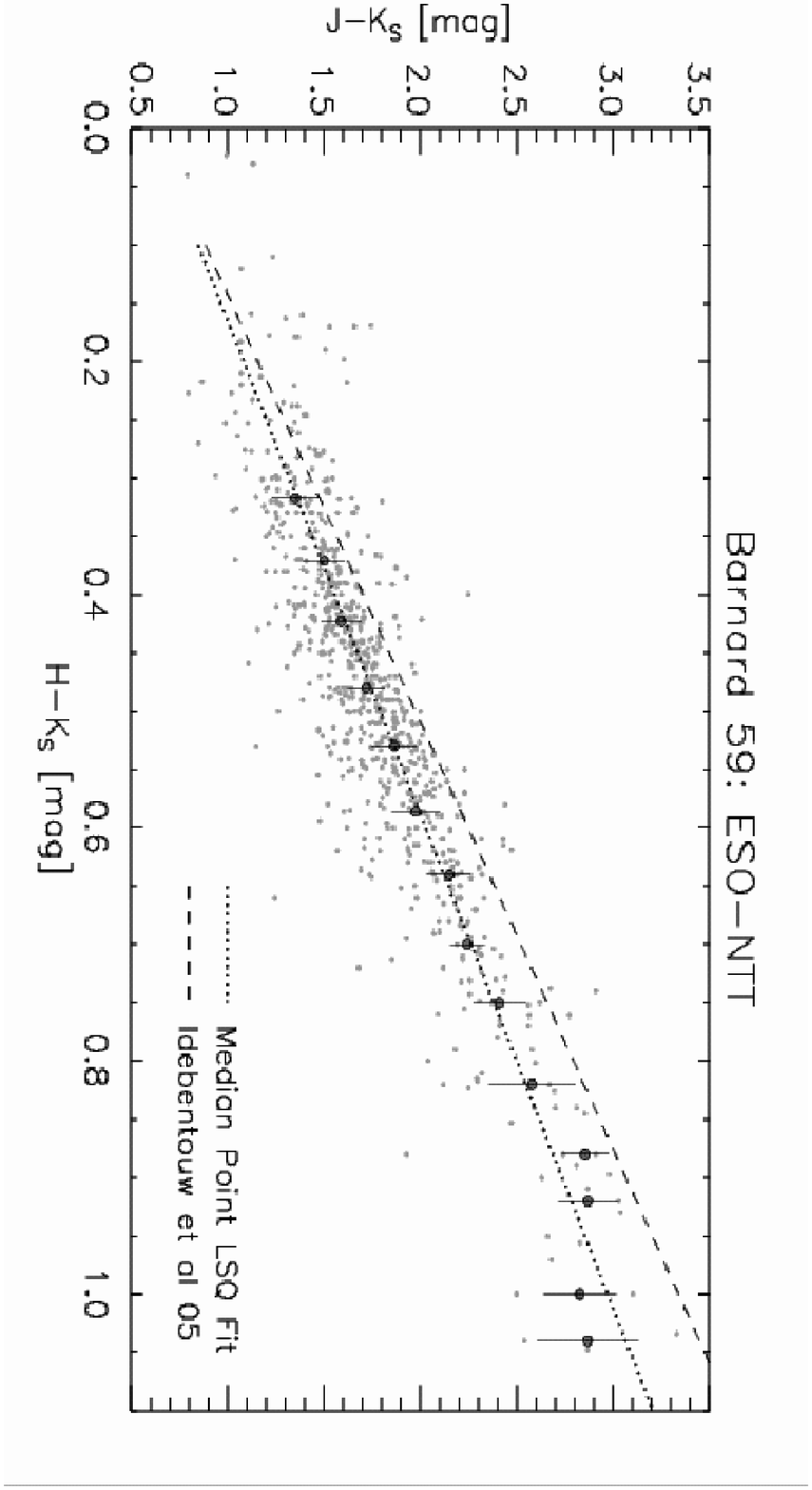}
\includegraphics[angle=90,scale=0.60]{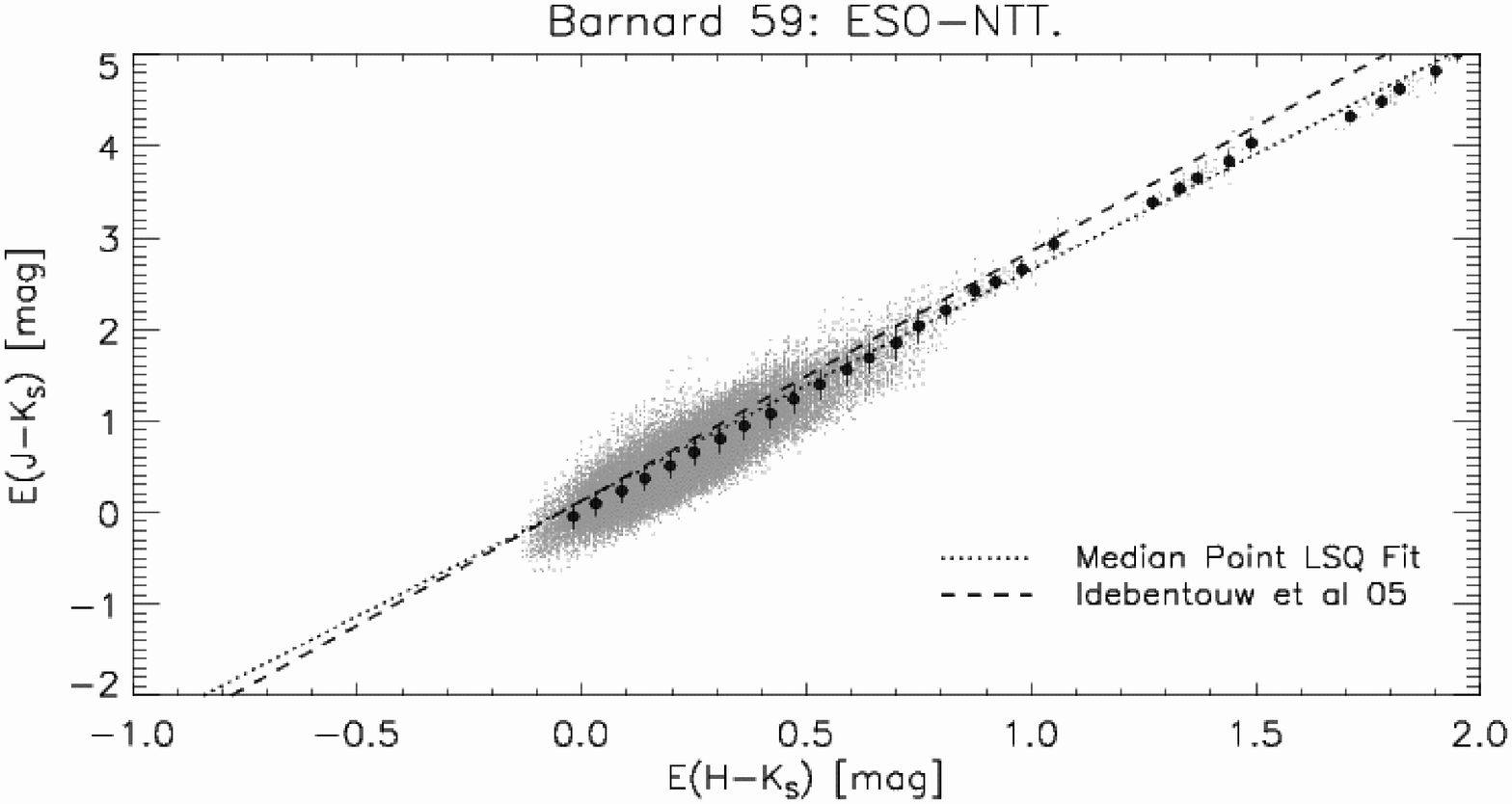}
\caption{$J-K$ vs. $H-K_S$ diagrams for sources with counterparts in both the Spitzer and ESO catalogs for the region of Barnard 59. {\it Top:} color-color diagram, {\it Bottom:} color-excess diagram (control sources are from an extinction-free field at the same galactic latitude as B59). In both plots gray points are ESO sources, the dotted line is the fit to the median points indicated by the large solid dots and the dashed line represents the median slope for sources in the fields studied by I05 -which has been shifted arbitrarily in the vertical direction to adjust to the ordinate of our fit.}
\label{fig:fit1}
\end{figure}

\begin{figure}
\includegraphics[angle=90,scale=0.32]{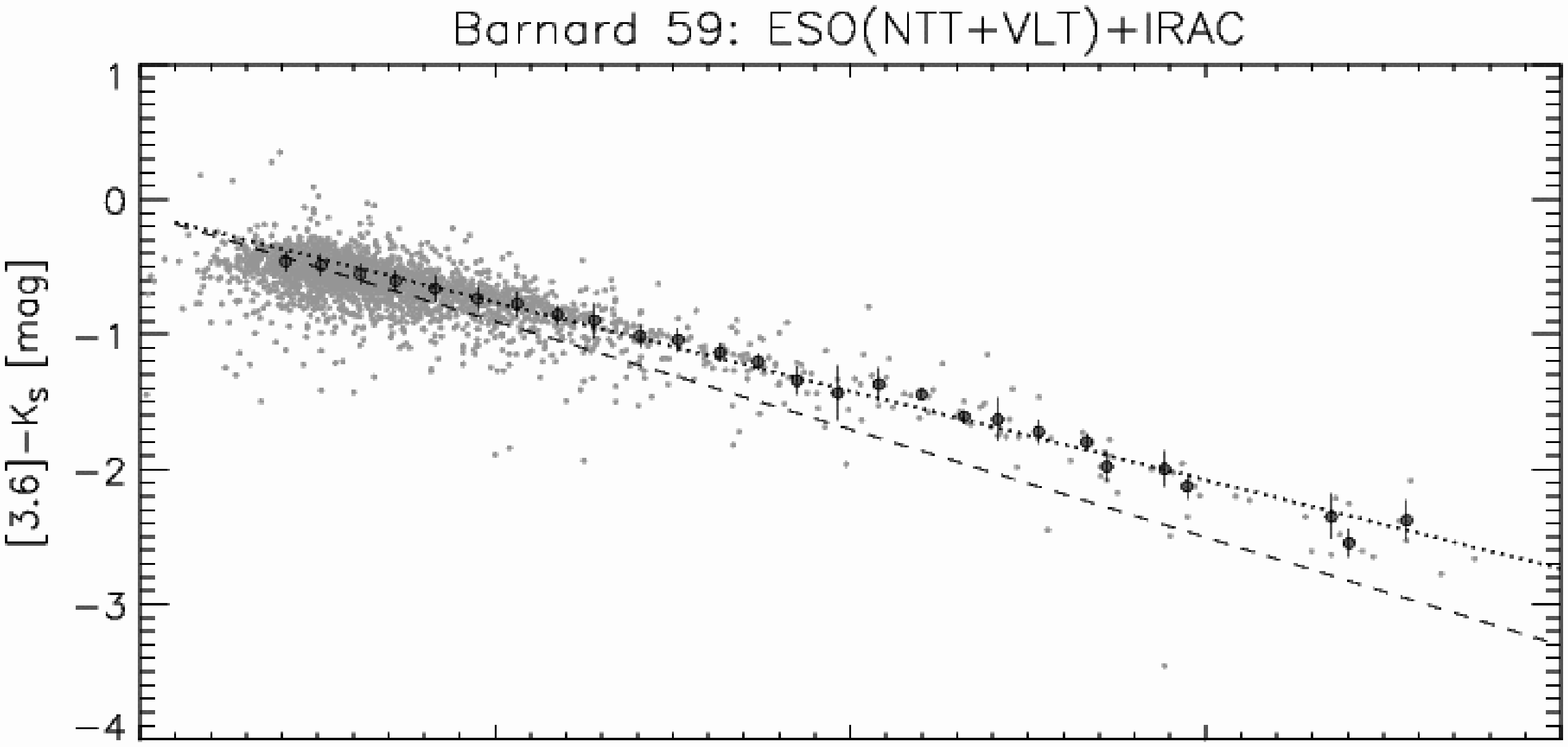}\includegraphics[angle=90,scale=0.32]{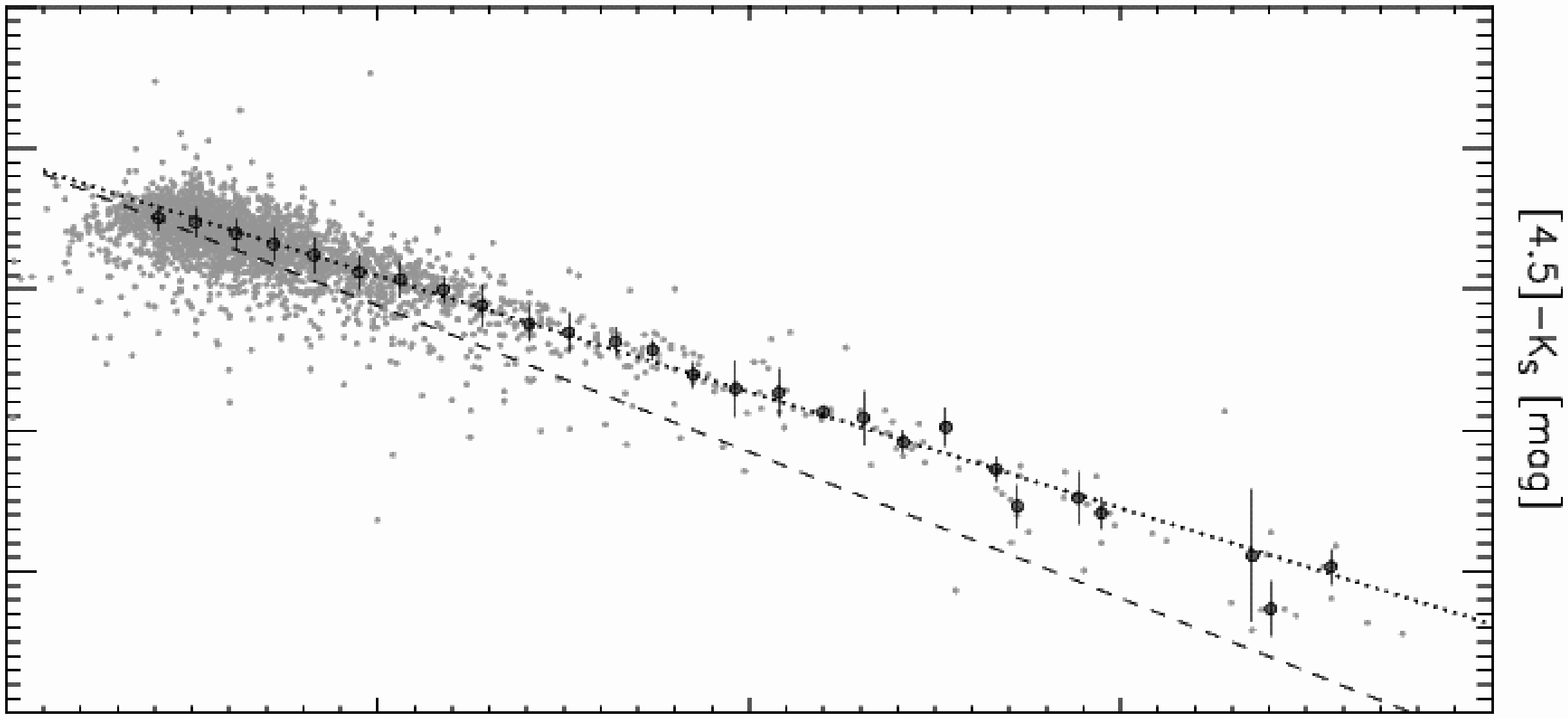}
\includegraphics[angle=90,scale=0.32]{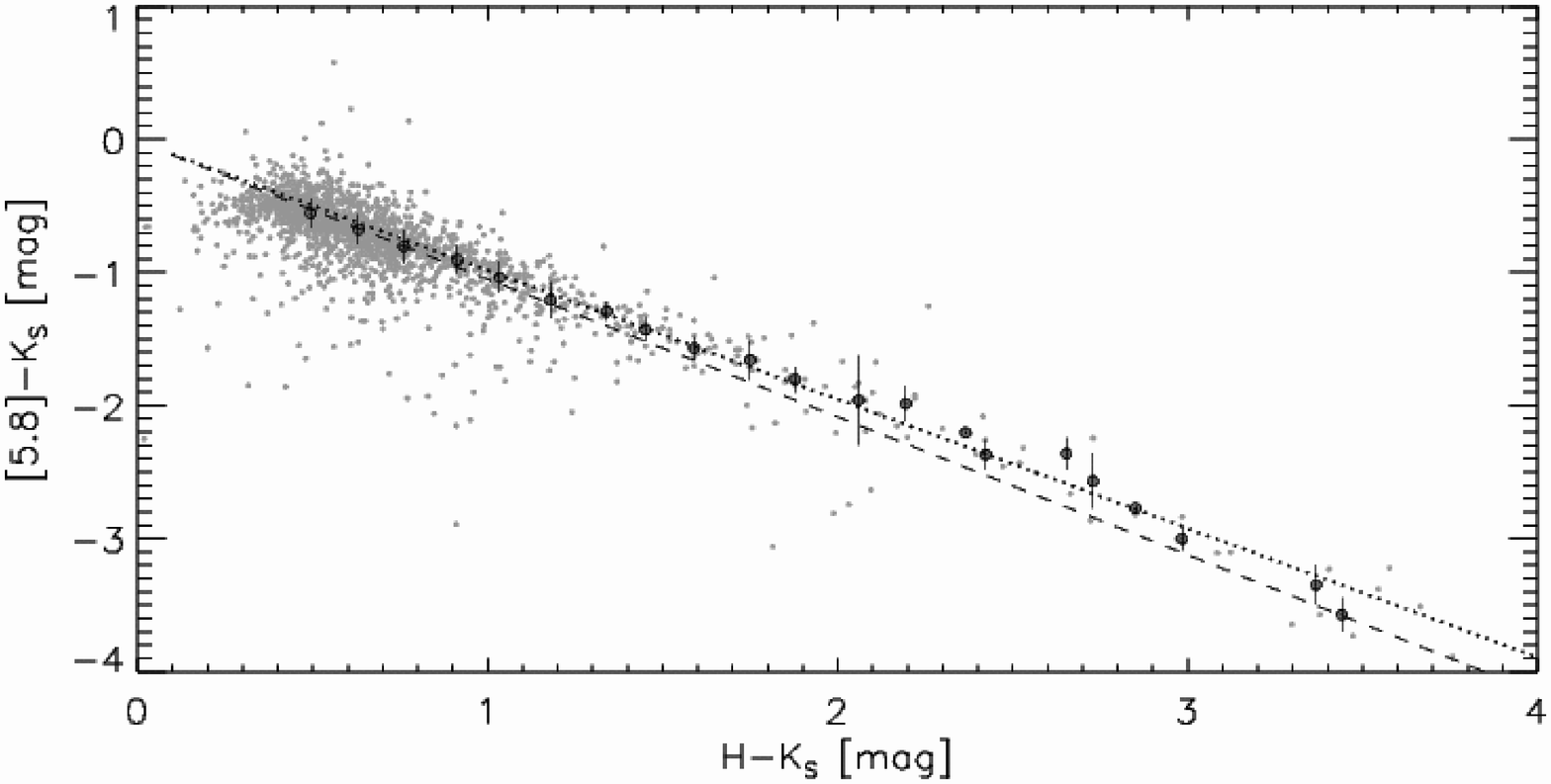}\includegraphics[angle=90,scale=0.32]{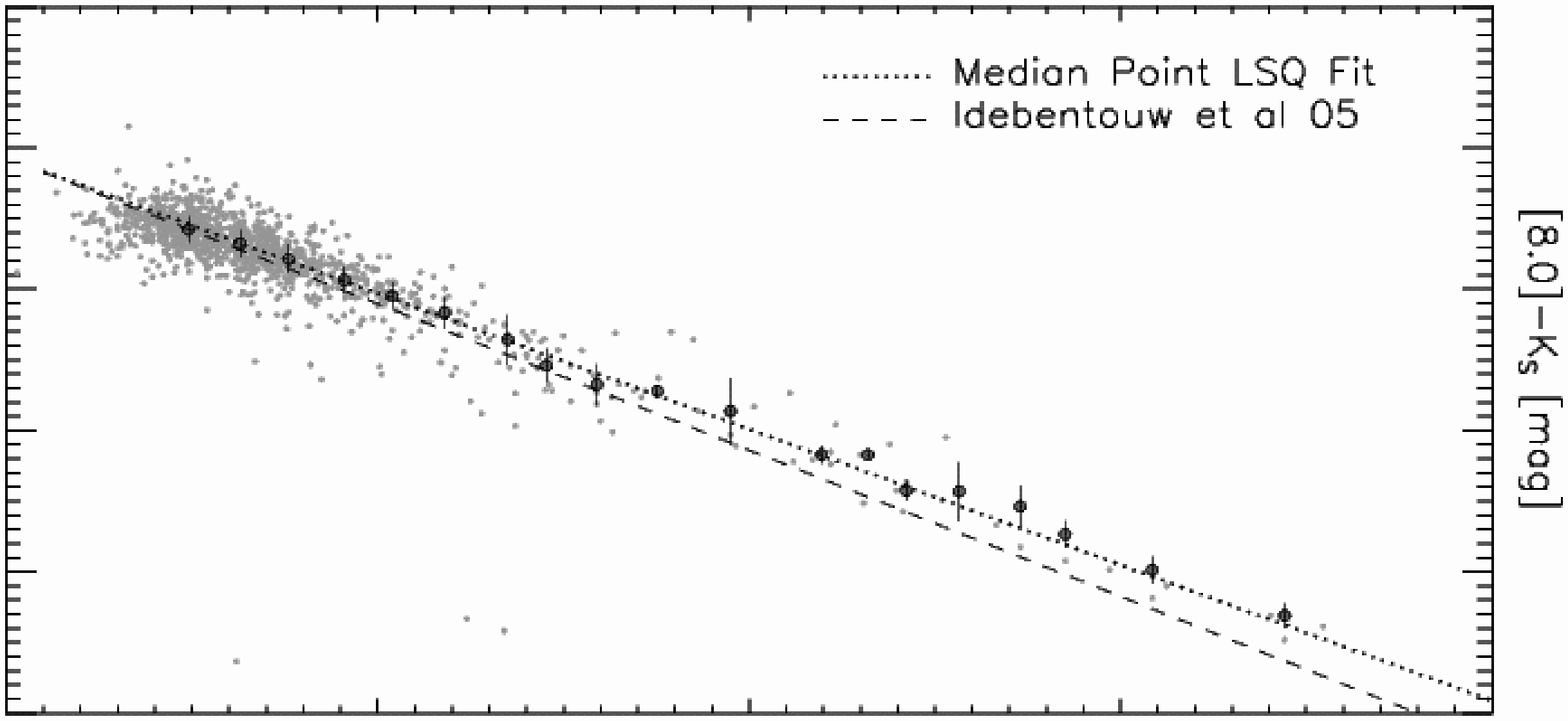}
\caption{Color-color diagrams for combined ESO/IRAC colors. Top left: $[3.6]-K$ vs. $H-K_S$ (R2); top right: $[4.5]-K$ vs. $H-K_S$ (R3); bottom left: $[5.8]-K$ vs. $H-K_S$ (R4); bottom right: $[3.6]-K$ vs. $H-K_S$ (R5). The layout of symbols and lines is the same as in Fig. \ref{fig:fit1}.}
\label{fig:fits2to5}
\end{figure}

The slopes of the distributions of B59 sources in our color-color diagrams are less steep than those inferred from the relative extinction ratios $A_\lambda /A_{K_S}$ calculated by I05 towards two distinct fields of the Galactic plane (see figures \ref{fig:fit1} and \ref{fig:fits2to5}). As a result of this, the extinction law, $A_\lambda/A_{K_S}$ as a function of $\lambda$, towards B59 is also less steep than the one determined by I05. \citet{ct06} derived an extinction curve in the form of a polynomial fit to the combined data points of I05 and L96, and we compare this curve to our data points in Figure \ref{fig:relative}. The difference we observe is modest, especially considering the large uncertainties in the relative extinction ratios of I05 and the data points of L96, but it could indicate that the extinction law in the NIR and MIR may vary slightly as a function of environment. In particular, I05 and L96 observations not only sample different lines of sight but also a not well constrained mixture of diffuse and dense cloud material, while our observations only sample dense cloud dust.

Our study was successful at combining NIR and MIR photometry to derive the extinction law through a highly opaque core in a molecular cloud. Barnard 59 is a good example of a cloud so obscured that NIR observations alone cannot penetrate it, and thus to detect and determine the colors of stars in its background, MIR observations are required. This situation is common to the centers of other very dense pre-stellar and young star forming cores, and even relatively thin IRDCs. In our experiment, the extinction law is measured specifically at opacity levels similar to those of such regions, and can be used in combination with deep NIR and MIR observations to probe highly obscured clouds with more confidence than if one had to use an extinction law extrapolated from measurements at lower opacities. It is thus possible to construct very deep extinction maps using Spitzer data, as we will show in an upcoming paper \citep{b59paper}, where we use Spitzer observations and our derived extinction law to construct an extinction map for B59 that reaches a depth close to $A_{K_S}=10$ mag, and resolves the most opaque regions near its center. Detailed knowledge of the inner structure of such a core can provide critical information about its physical state \citep{b68,kandori,lupus}, and its stability against gravitational collapse \citep{shu77,coalsack,harvey}. 

\acknowledgments

We are grateful to Thomas Dame and Tracy Huard for useful discussions. This work is based on observations made with ESO Telescopes at the La Silla or Paranal Observatories under programme ID 069.C-0426. This work is also based on observations made with the Spitzer Space Telescope, which is operated by the Jet Propulsion Laboratory, California Institute of Technology, under NASA contract 1407. We acknowledge support from the NASA's {\it Origins} Program (NAG 13041) and the Spitzer grant GO20119. This publication makes use of data products from the Two Micron All Sky Survey, which is a joint project of the University of Massachusetts and the Infrared Processing and Analysis Center/California Institute of Technology, funded by the National Aeronautics and Space Administration and the National Science Foundation. 

{\it Facilities:} \facility{NTT (SOFI)}, \facility{VLT (ISAAC)}, \facility{Spitzer (IRAC)}.

\clearpage

\clearpage

\clearpage

\clearpage

\clearpage

\clearpage

\begin{deluxetable}{ccccccc}
\tablecolumns{7} 
\tablewidth{0pc} 
\tablecaption{The Extinction Law between 1.2 and 8.0~$\mu \rm m$ in Barnard 59 \label{tab:extlaw}} 
\tablehead{
\colhead{$\lambda$ \tablenotemark{a}} &
\multicolumn{3}{c}{$E(\lambda-K_S)/E(H-K_S)$} &
\colhead{} &
\multicolumn{2}{c}{$A_\lambda/A_{K_S}$} \\
\cline{2-4} \cline{6-7}
\colhead{[$\mu$m]} &
\colhead{I05 \tablenotemark{b}} &
\colhead{C-C \tablenotemark{c}} &
\colhead{C-E \tablenotemark{d}} &
\colhead{} &
\colhead{C-C} &
\colhead{C-E} \\
}
 
\startdata
1.240  &  2.73 &  2.36$\pm$0.18 &  2.52$\pm$0.05 &  &  2.299$\pm$0.530 &  2.389$\pm$0.045 \\
1.664  &  1.00 &  1.00$\pm$0.00 &  1.00$\pm$0.00 &  &  1.550$\pm$0.080 &  1.550$\pm$0.080 \\
2.164  &  0.00 &  0.00$\pm$0.00 &  0.00$\pm$0.00 &  &  1.000$\pm$0.000 &  1.000$\pm$0.000 \\
3.545  & -0.80 & -0.66$\pm$0.02 & -0.69$\pm$0.03 &  &  0.618$\pm$0.077 &  0.619$\pm$0.029 \\
4.442  & -1.04 & -0.83$\pm$0.03 & -0.89$\pm$0.04 &  &  0.525$\pm$0.063 &  0.512$\pm$0.022 \\
5.675  & -1.04 & -0.97$\pm$0.02 & -0.98$\pm$0.05 &  &  0.462$\pm$0.055 &  0.459$\pm$0.024 \\
7.760  & -1.04 & -0.96$\pm$0.02 & -0.97$\pm$0.04 &  &  0.455$\pm$0.059 &  0.465$\pm$0.017 \\
\enddata

\tablenotetext{a}{Isophotal wavelengths of 2MASS and IRAC convolved with a K2 III star (see I05).}
\tablenotetext{b}{Slopes of I05 inferred from the $A_\lambda/A_{K_S}$ ratios of their Table 1.}
\tablenotetext{c}{Slopes of the distributions of source colors.}
\tablenotetext{d}{Slopes of the distributions of source color-excesses.}

\end{deluxetable} 

\end{document}